**Title**

Identifying heat-related diagnoses in emergency department visits among adults in Chicago: a heat-wide association study

**Authors**
Hyojung Jang,[1]* Peter M. Graffy,[1]† Benjamin W. Barrett,[1]† Daniel E. Horton,[2] Jennifer L. Chan,[3] Abel N. Kho[1, 4]

*Corresponding author, Email: hyojung.jang@northwestern.edu
†These authors contributed equally to this work

**Affiliations**

1. Division of Biostatistics & Informatics, Department of Preventive Medicine, Feinberg School of Medicine, Northwestern University, Chicago, IL, 60611, USA.

2. Department of Earth, Environmental, and Planetary Sciences, Northwestern University, Evanston, IL, 60208, USA.

3. Department of Emergency Medicine, Feinberg School of Medicine, Northwestern University, Chicago, IL, 60611, USA.

4. Department of Medicine, Feinberg School of Medicine, Northwestern University, Chicago, IL, 60611, USA.

**Abstract**

Extreme heat is an escalating public health concern. Although prior studies have examined heat–health associations, their reliance on restricted diagnoses and diagnostic categories misses or misclassifies heat-related illness. We conducted a heat-wide association study to identify acute-care diagnoses associated with extreme heat in Chicago, Illinois. Using 916,904 acute-care visits—including emergency department and urgent care encounters—among 372,140 adults across five healthcare systems from 2011–2023, we applied a two-stage analytic approach: quasi-Poisson regression to screen 1,803 diagnosis codes for heat-related risks, followed by distributed lag non-linear models in a time-stratified case-crossover design to refine the list of heat-related diagnoses and estimate same-day and short-term cumulative odds ratios of acute-care visits during extreme heat versus reference temperature. We observed same-day increases in visits for heat illness, volume depletion, hypotension, edema, acute kidney failure, and multiple injuries. By analyzing the full diagnostic spectrum of acute-care services, this study comprehensively characterizes heat-associated morbidity, reinforcing and advancing existing literature.

**MAIN TEXT**

**Introduction**

Extreme heat exposure is an escalating public health concern, with climate change driving increases in the frequency, intensity, and duration of heat events (1, 2). Both the frequency and duration of heat events have increased substantially since the 1950s, with nearly one-third of the global population now experiencing at least 20 days of extreme heat annually (3, 4). Alongside this global trend, urban environments face additional risks due to the urban heat island (UHI) effect, which can elevate local temperatures by several degrees and prolong heat events (5, 6).

Extreme heat places a substantial strain on health systems, prompting extensive study of its effects on morbidity, mortality, and healthcare utilization. Studies have demonstrated that extreme heat is associated with increased all-cause mortality (7-15) and heat-specific illnesses such as heat exhaustion and heat stroke (16-18). Morbidity also rises during extreme heat, including cardiovascular and respiratory events (19-22), renal disease and electrolyte disturbances including hyponatremia (19, 23-26), worsening mental health outcomes (27, 28), and increased injury incidence (7, 29). Consistent with these clinical impacts, healthcare use increases, with surges in emergency department (ED) visits and hospital admissions (16-19, 22-25, 30-35).

Despite these advances in our understanding of the health effects of extreme heat, most studies have taken a deductive, hypothesis-driven approach with pre-selected diagnoses and diagnostic categories to define heat-related illness, yielding valuable but incomplete coverage of heat-related morbidity. Furthermore, heat-related health outcomes are difficult to define due to their diverse symptom presentation leading to non specific documentation of conditions or diagnoses, and undercategorization of clinical conditions as explicitly "heat-related"—especially in emergency settings (36). These challenges complicate studies on heat-associated morbidity and can lead to underestimation of the true public health impact of extreme heat. Moreover, such underestimation may obscure the disproportionate impacts experienced by socially and economically vulnerable populations, who often have limited adaptive capacity and face greater risks from extrem heat (37-39). Addressing this knowledge gap is critical to more completely capture the full extent of heat-related health risks, which can inform preparedness strategies and support equitable public health interventions.

Here, we introduce a 'heat-wide association study' (HWAS) – an inductive, data-driven framework to systematically evaluate the short-term impact of extreme heat across the full spectrum of International Classification of Disease (ICD)-coded diagnoses in ED visits within a large integrated urban healthcare network in Chicago, Illinois (40). Chicago—marked by high-density development, a pronounced UHI, and a documented history of heat-related morbidity and mortality—provides the sociodemographic diversity and healthcare infrastructure needed to study ED-based heat risks at the city level (9, 33, 41, 42). The Chicago Area Patient-Centered Outcomes Research Network (CAPriCORN) (43) links electronic health record (EHR) data from multiple Chicago-based healthcare systems, representing a sociodemographically and geographically diverse urban patient population. Patient residential data in CAPriCORN enables precise linkage of environmental exposures to clinical outcomes, providing a robust platform for studying conditions with increased acute-care visit risk during extreme heat events (44). Our analysis uses retrospective acute-care encounters—emergency department (ED) and urgent care (UC) visits—from five healthcare systems in CAPriCORN. Throughout this manuscript, we use emergency department (ED) as a general term encompassing both ED and urgent care (UC) visits, collectively referred to as acute-care services.

The analytic cohort comprised 916,904 ED visits among 372,140 unique adult ($\geq$ 18 years) patients during the warm-season (May-September) from 2011 through 2023. We examined all 1,803 distinct ICD-10 and ICD-9 equivalent diagnosis categories, as captured in CAPriCORN ED visits, as outcomes of interest. We standardized ICD-9 codes to their corresponding ICD-10 categories for consistency (see Methods). We implemented a two-stage analytic framework—quasi-Poisson screening followed by time-stratified case-crossover distributed lag non-linear models (DLNMs)—to efficiently screen and quantify associations between extreme heat and ICD-10 diagnostic categories. Figure 1 describes the overall analytic approach. We stratified analyses by demographic and neighborhood characteristics to identify disparities in heat vulnerability. We also conducted sensitivity analyses to evaluate the robustness of our model specifications and findings. Leveraging more than a decade of multi-institutional EHR data integrated with high-resolution temperature data, this study provides a robust and comprehensive evaluation of extreme heat-related diagnoses at a citywide scale. This approach not only enables the identification of underrecognized heat-sensitive conditions, but also facilitates subgroup analyses to uncover heat-related health outcome disparities across demographic and neighborhood contexts (see Methods for analytic details).

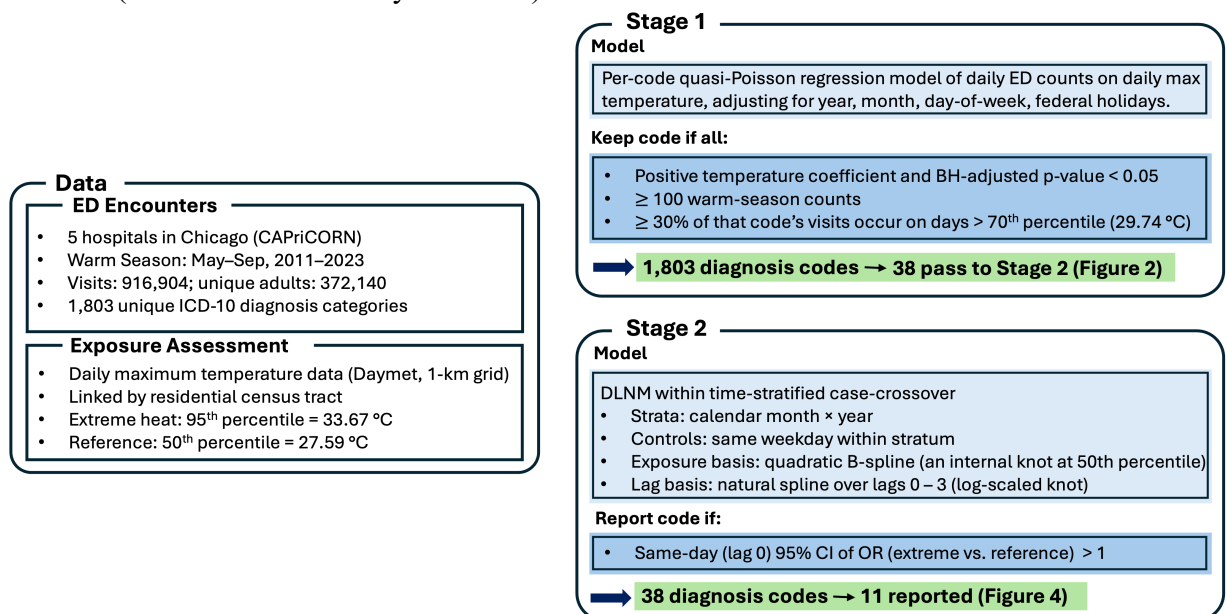

Figure 1. Two-stage analytic workflow. Stage 1 screened 1,803 ICD-10 diagnosis codes with per-code quasi-Poisson models of daily ED visit counts versus daily maximum temperature; codes with a positive slope, BH-adjusted P < 0.05, ≥ 100 warm-season counts, and ≥ 30% of visits on > 70th-percentile temperature days were retained (n = 38). Stage 2 fit distributed lag nonlinear models (DLNM) in a time-stratified case-crossover (lags 0–3) to estimate odds ratios for the 95th vs 50th temperature percentile; codes with lag-0 lower 95% CI > 1 were reported (n = 11). BH = Benjamini–Hochberg; CAPriCORN = Chicago Area Patient-Centered Outcomes Research Network; CI = Confidence interval; ED = Emergency department; OR = Odds ratio.

## Results

### Study population and heat exposure summary

Among 372,140 adults with ED visits, females accounted for 57.9% of patients, and the mean age of patients was 52 years (standard deviation, 19.4 years). Table 1 summarizes patient demographics across ED visits and the geographic distributions of residences by heat exposure category. Over the study period, there were 100 extreme heat days ($\geq 33.67°C$ [92.60°F]; 95$^{th}$ percentile of daily maximum temperature during the warm-season) and 1,882 non-extreme heat days. Extreme heat days were concentrated in mid-summer, occurring most frequently in July (36%), followed by August (25%), June (23%), September (12%), and May (4%). When stratified by year, the highest frequency of extreme heat days was observed in 2012 (25%), followed by 2018 (15%), and 2011 (13%).

Demographic and geographic distributions were broadly similar between non–extreme heat and extreme heat days, indicating comparable patient profiles across exposure categories. Adults aged 25–44 years represented the largest age group, Black or African American patients comprised the majority by race/ethnicity, and the West Side of Chicago contributed the highest share of ED visits.

|  | Non-extreme heat days (N = 884,662) | Extreme heat days (N = 32,242) | Total (N = 916,904) |
|---|---|---|---|
| **Age (years)** | | | |
| 18-24 | 90,496 (10.2%) | 3,541 (11.0%) | 94,037 (10.3%) |
| 25-44 | 335,891 (38.0%) | 12,404 (38.5%) | 348,295 (38.0%) |
| 45-64 | 257,275 (29.1%) | 9,276 (28.8%) | 266,441 (29.1%) |
| 65 + | 156,279 (17.7%) | 5,433 (16.9%) | 161,712 (17.6%) |
| Missing | 44,721 (5.1%) | 1,588 (4.9%) | 46,309 (5.1%) |
| **Sex** | | | |
| Female | 511,695 (57.8%) | 18,683 (57.9%) | 530,378 (57.8%) |
| Male | 372,804 (42.1%) | 13,553 (42.0%) | 386,357 (42.1%) |
| Other | 60 (0.0%) | 2 (0.0%) | 62 (0.0%) |
| Missing | 103 (0.0%) | 4 (0.0%) | 107 (0.0%) |
| **Race/Ethnicity** | | | |
| Asian | 15,945 (1.8%) | 577 (1.8%) | 16,522 (1.8%) |
| Black or African American | 472,467 (53.4%) | 17,655 (54.8%) | 490,122 (53.5%) |
| White | 244,368 (27.6%) | 8,984 (27.9%) | 253,352 (27.6%) |
| Other | 114,822 (13.0%) | 3,699 (11.5%) | 118,521 (12.9%) |
| Missing | 37,060 (4.2%) | 1,327 (4.1%) | 38,387 (4.2%) |
| **Region of Chicago** | | | |
| Central Side | 92,968 (10.5%) | 3,077 (9.5%) | 96,045 (10.5%) |
| Far South Side | 55,316 (6.3%) | 1,895 (5.9%) | 57,211 (6.2%) |
| North Side | 85,679 (9.7%) | 2,971 (9.2%) | 88,650 (9.7%) |
| Northwest Side | 74,368 (8.4%) | 2,618 (8.1%) | 76,986 (8.4%) |
| South Side | 101,818 (11.5%) | 3,666 (11.4%) | 105,484 (11.5%) |
| Southwest Side | 118,012 (13.3%) | 4,593 (14.2%) | 122,605 (13.4%) |
| West Side | 275,639 (31.2%) | 10,546 (32.7%) | 286,185 (31.2%) |
| Missing | 80,862 (9.1%) | 2,876 (8.9%) | 83,738 (9.1%) |

**Table 1. Cohort characteristics by heat exposure category.** Characteristics of patients across emergency department visits by heat exposure category during warm-season months (May-September), 2011-2023. Non-extreme heat days are dates with daily maximum temperature < $95^{th}$ percentile; extreme heat days are ≥ $95^{th}$ percentile of the warm-season daily maximum temperature distribution. Age (18–24, 25–44, 45–64, ≥ 65 years), sex (Female, Male, Other), and race/ethnicity (Asian, Black or African American, White, Other) are reported as recorded in the electronic health record; "Missing" reflects unavailable data. Geographic region corresponds to Chicago community regions (Central, North, Northwest, South, Southwest, Far South, West Side), assigned using residential census tract centroids.

**Analytic framework overview**

We applied a two-stage analytic framework to systematically identify and quantify diagnosis-specific associations between extreme heat and ED visits. In the first stage, we fit diagnosis-specific quasi-Poisson regressions of daily ED visit counts on daily maximum temperature to identify codes whose counts increased on hotter days, controlling the false discovery rate with Benjamini–Hochberg (BH) adjustment (45). In the second stage, we fit DLNMs in a time-stratified case-crossover design (13, 46) to estimate the immediate (same-day) and short-term cumulative associations between extreme heat and diagnosis codes identified in the first stage. The two-stage analysis enhances efficiency and rigor by containing the multiple-testing burden in the first stage while preserving detailed exposure quantification through DLNMs in the second stage.

**First-stage screening analysis of diagnosis codes**

The first-stage screening using quasi-Poisson modeling identified 38 ICD-10 diagnosis codes that met three criteria: (1) a statistically significant positive association between daily ED visit counts for the diagnosis during warm-season and daily maximum temperature after BH adjustment, (2) ≥ 30% of ED visits for the diagnosis occurred on days when the maximum temperature exceeded the 70th percentile (29.74°C [85.53°F]) of the distribution, and (3) at least

100 counts of the diagnosis code recorded during warm-season in the ED systems. Figure 2 shows the first-stage screening results in a Manhattan plot. Each point represents an individual diagnosis code, with the vertical axis denoting statistical significance ($-\log_{10}$ of BH-adjusted p-value) from quasi-Poisson regression of daily ED visit counts on same-day maximum temperature. The dashed red line marks the multiple-testing significance threshold (adjusted p-value < 0.05). Labels identify codes that also met the frequency criteria (≥ 30% of ED visits on days above the 70th percentile [29.74°C] and ≥ 100 counts of diagnosis code). This representation highlights well-recognized diagnoses associated with extreme heat, such as heat stroke (T67), exposure to excessive natural heat (X30), and dehydration (E86), while also revealing additional conditions spanning circulatory, renal, dermatologic, and injury-related categories.

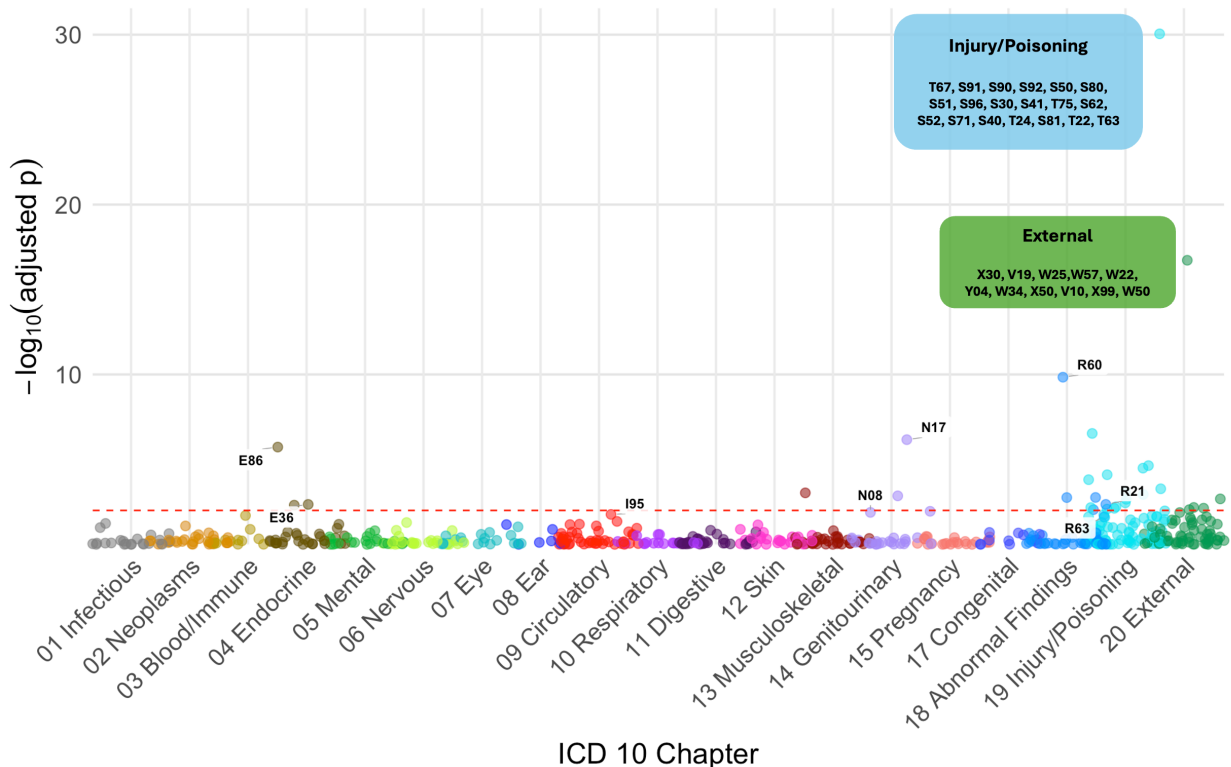

**Figure 2. Manhattan plot of diagnosis codes associated with daily maximum temperature.** Each dot represents an ICD-10 diagnosis code, with the y-axis showing statistical significance from quasi-Poisson regression models (–log₁₀ Benjamini–Hochberg-adjusted p-value). The red dashed line marks the multiple-testing threshold (Benjamini–Hochberg-adjusted p < 0.05). Text labels are displayed for all codes that met all screening criteria: (1) a significant positive association between emergency department visit counts and daily maximum temperature, (2) ≥ 30% of emergency department visits occurring on days above the 70th percentile of the temperature distribution (29.74°C), and (3) at least 100 counts of the diagnosis code recorded during warm-season (May-September) in the emergency department systems. Because two chapters contained many codes that met all screening criteria—Injury/Poisoning and External causes—their labels are placed in inset boxes to avoid overplotting. The insets list **all** passing codes from those chapters, ordered by increasing BH-adjusted p-value (smallest to largest); the corresponding points remain in their chapter positions in the main panel.

Figure 3 displays the 38 diagnosis codes identified in the first-stage screening, stratified into (a) disease-related outcomes, and (b) injury, poisoning, and external causes. Points indicate incidence rate ratios (IRRs) with 95% confidence intervals (CIs), quantifying the relative change in ED visit rates per 1°C increase in daily maximum temperature; the dashed line denotes IRR = 1 (no association). A complete listing of diagnoses, including ICD-10 descriptions, chapters, IRRs, BH-adjusted p-values, and the proportion of visits occurring on days above the 70th percentile of the temperature distribution, is provided in Supplementary Table S1.

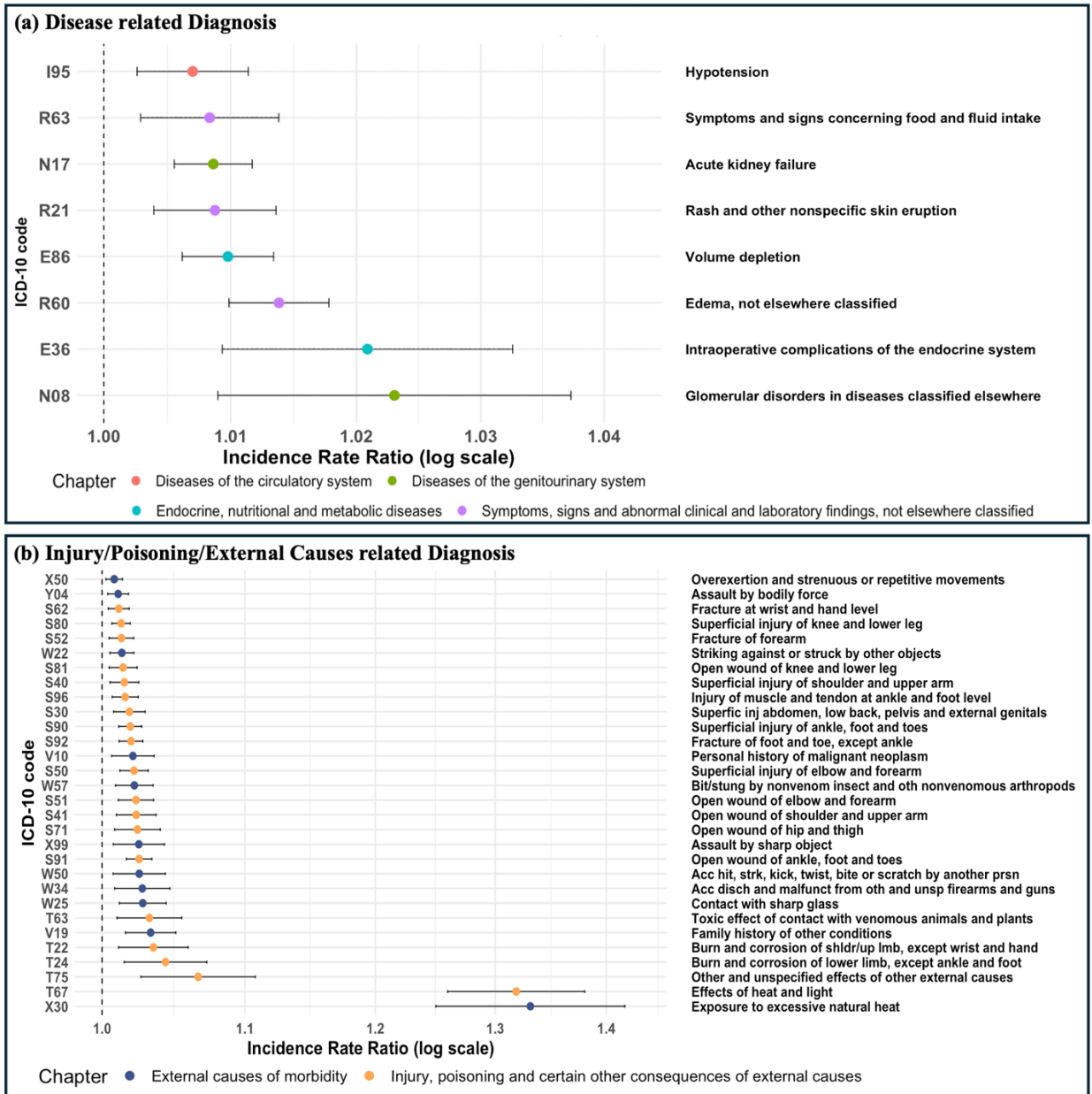

**Figure 3. Screening results of heat-sensitive emergency department (ED) diagnoses.** Points show incidence rate ratios (IRR) from quasi-Poisson models relating daily maximum temperature to ED visit counts by diagnosis codes; horizontal bars are 95% confidence intervals (CIs). The top panel (a) displays disease-related (non-injury) ICD-10 codes; the bottom panel (b) shows injury/poisoning/external-cause ICD-10 codes. Different colors indicate ICD-10 chapter; the dashed vertical line marks IRR = 1 (no association). Diagnoses in the plot met the a priori screening criteria (Benjamini–Hochberg adjusted P-value < 0.05, ≥ 30% of ED visits occurring on days above the 70th percentile of the warm-season (May-September) daily maximum temperature distribution, and at least 100 counts of the diagnosis code recorded during warm-season in the ED systems).

## Second-stage lag-specific and cumulative effects of extreme heat

For the second-stage of our analysis, we restricted the evaluation to the 38 ICD-10 diagnosis codes that passed the first-stage screening, and estimated both immediate and short-term temperature effects using DLNMs within a time-stratified case-crossover design. By focusing on the subset of diagnoses with evidence of temperature sensitivity, we enhanced interpretability while reducing the likelihood of spurious associations. The DLNM framework within the case-crossover design allowed us to capture non-linear exposure–response relationships over lag 0–3 days.

Figure 4 highlights the subset of diagnosis codes whose 95% CI of same-day odds ratio (OR) of ED visits at extreme heat (≥ 33.67°C [92.60°F]) relative to the median warm-season temperature (27.59°C [81.66°F]) lies entirely above 1, indicating a higher likelihood of ED visits under extreme heat conditions compared with the median temperature. We identified 11 diagnosis codes meeting the criterion of elevated same-day risk. Among these, the largest risks were for explicitly heat-related codes: **Exposure to excessive natural heat** (X30; OR = 6.65, 95% CI: 3.43-12.87) and **Effects of heat and light** (T67; OR = 6.15, 95% CI: 4.33-8.72). Beyond the explicit heat-related conditions, elevated risks were also observed for several injury and external-causes categories, including **Foot/ankle injuries** (S91, S96), **Accidental discharge and malfunction from other firearms and guns** (W34), **Assault by bodily force** (Y04), and **Other and unspecified effects of other external causes** (T75). Significant immediate heat effects were also detected for multiple medical conditions: **Volume depletion** (loss of body fluids) (E86; OR = 1.12, 95% CI: 1.07–1.18), **Hypotension** (low blood pressure) (I95; OR = 1.09, 95% CI: 1.02–1.16), **Edema** (swelling) (R60; OR = 1.07, 95% CI: 1.00–1.14), and **Acute kidney failure** (N17; OR = 1.05, 95% CI: 1.00–1.10). Table 2 summarizes immediate (lag 0) and cumulative (lags 0–1, 0–2, and 0–3) ORs for the diagnoses shown in Figure 4, comparing extreme heat (33.67°C [92.60°F]) with the median temperature (27.59°C [81.66°F]).

**(a) Disease related Diagnosis**

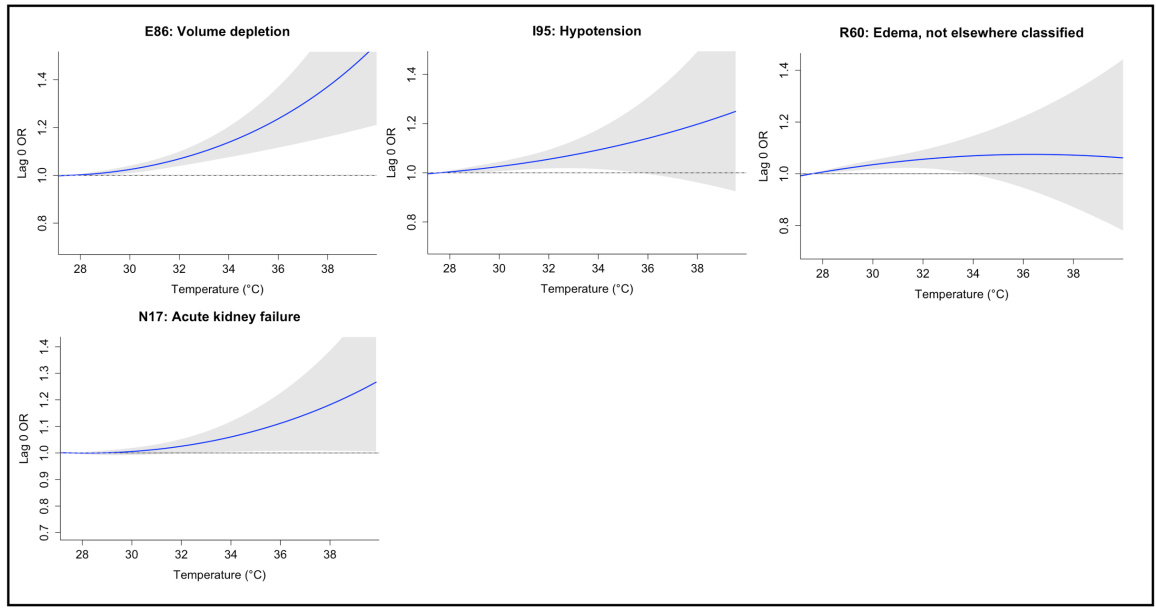

**(b) Injury/Poisoning/External Causes related Diagnosis**

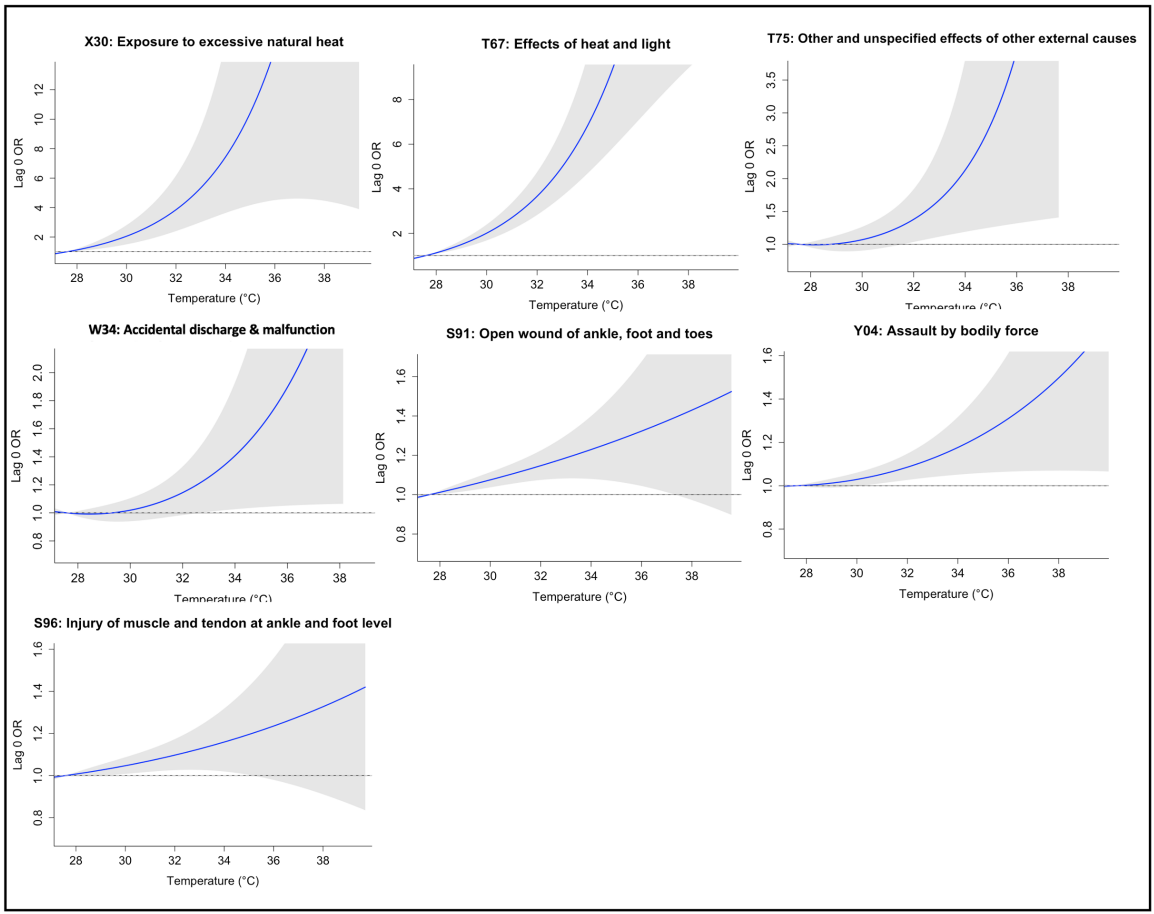

**Figure 4. Immediate (same day, lag 0) heat effects on emergency department (ED) visits by diagnosis.** Immediate temperature–ED associations for ICD-10 codes retained after first-stage screening, estimated with distributed lag non-linear models in a time-stratified case-crossover design. Curves show odds ratios (ORs) across extreme heat temperature (33.67°C) relative to the median (27.59°C); shaded bands are 95% confidence intervals (CIs) and the horizontal dashed line marks OR = 1. Panels: (a) disease diagnoses; (b) injury/poisoning/external-cause diagnoses. Only codes with a 95% CI lower bound > 1 at the 95th percentile temperature (33.67°C) are displayed.

|  | Same day OR (lag 0) | Cumulative OR (lag 0-1) | Cumulative OR (lag 0-2) | Cumulative OR (lag 0-3) |
|---|---|---|---|---|
| **Disease-related Diagnosis** | | | | |
| **E86:** Volume depletion | 1.12 (1.07–1.18) | 1.14 (1.05–1.24) | 1.12 (0.99–1.26) | 1.10 (0.95–1.27) |
| **I95:** Hypotension | 1.09 (1.02–1.16) | 1.14 (1.03–1.27) | 1.14 (0.98–1.32) | 1.07 (0.89–1.29) |
| **R60:** Edema | 1.07 (1.00–1.14) | 1.13 (1.02–1.25) | 1.17 (1.02–1.34) | 1.18 (1.00–1.40) |
| **N17:** Acute kidney failure | 1.05 (1.00–1.10) | 1.09 (1.01–1.17) | 1.12 (1.00–1.24) | 1.16 (1.02–1.32) |
| **Injury/Poisoning/External Diagnosis** | | | | |
| **X30:** Exposure to excessive natural heat | 6.65 (3.43–12.87) | 4.92 (1.61–15.03) | 2.10 (0.40–10.89) | 1.60 (0.20–13.18) |
| **T67:** Effects of heat and light | 6.15 (4.33–8.72) | 7.75 (4.34–13.86) | 6.07 (2.68–13.75) | 6.34 (2.27–17.68) |
| **T75:** Other and unspecified effects of other external causes | 1.96 (1.17–3.29) | 2.60 (1.12–6.06) | 2.90 (0.92–9.08) | 3.10 (0.79–12.24) |
| **W34:** Accidental discharge and malfunction from other firearms and guns | 1.35 (1.02–1.79) | 1.49 (0.93–2.38) | 1.60 (0.82–3.14) | 1.91 (0.81–4.49) |
| **S91:** Open wound of ankle, foot and toes | 1.21 (1.08–1.36) | 1.37 (1.14–1.66) | 1.50 (1.15–1.95) | 1.62 (1.16–2.27) |
| **Y04:** Assault by bodily force | 1.16 (1.05–1.28) | 1.18 (1.00–1.39) | 1.18 (0.94–1.49) | 1.28 (0.96–1.70) |
| **S96:** Injury of muscle and tendon at ankle and foot level | 1.15 (1.02–1.29) | 1.23 (1.01–1.49) | 1.25 (0.95–1.64) | 1.21 (0.86–1.71) |

**Table 2.** Same day (lag 0) and cumulative (lags 0–1, 0–2, and 0–3) odds ratios (OR, 95% confidence interval [CI]) for the association between extreme heat and emergency department visits by ICD-10 code, estimated with distributed lag non-linear models in a time-stratified case-crossover design. Diagnoses shown were retained by first-stage screening and had 95% CI lower bounds > 1 for the same day effect. Codes are ordered by decreasing same day OR and grouped as disease-related diagnoses and injury/poisoning/external-cause diagnoses.

**Heat vulnerability across demographic and geographic strata**

As a diagnostic step, we first compared baseline demographic (age, sex, race/ethnicity) and geographic (residential region) distributions of ED visits for each of the 11 diagnosis codes identified in the second-stage analysis, stratified by non-extreme and extreme heat days. Supplementary Materials (Figures S1-S4) visualize these distributions. Then, to assess potential effect modification, we re-estimated the time-stratified case-crossover DLNM for each diagnosis captured in the first-stage screening analysis, stratifying by sex, age group, race/ethnicity, and geographic region. We obtained immediate ORs and cumulative ORs over lag 0-3 days within each stratum.

Across age strata, ***Volume depletion*** (E86) was consistently elevated in young (18-24 years), middle-aged (25-44 years), and late middle-aged adults (45-64 years). Elevated odds of ankle and foot injuries were also consistently observed across multiple adult age groups (25 years and older). Other diagnoses displayed age-specific patterns; ICD-10 codes meeting the criterion of a lag-0 OR with a lower CI bound greater than 1 across age strata are described in Table 3. Among the 18-24 years age group, heat was strongly associated with ***Rash and other nonspecific skin eruption*** (R21). In the 25-44 years age group, associations concentrated in ***Upper- and lower-limb trauma***, including shoulder/upper-arm (S40) and ankle/foot injuries (S96, S91, S90), with additional signals for ***Accidental discharge and malfunction from firearms and guns*** (W34). Among individuals aged 45–64 years, ***Fractures of foot and toe*** (S92) remained prominent, while ***Edema*** (R60) was also observed. Among the ≥ 65 years age group, the diagnosis code set narrowed to ***Fracture of foot and toe*** (S92) and ***Hypotension*** (I95). Because of sparse events, estimates for ***Exposure to excessive natural heat*** (X30) were unstable in all age strata and are not tabulated. For ***Effects of heat and light*** (T67), only the 45–64 years stratum yielded a stable estimate (OR 4.57, 95% CI 2.46–8.50); all other strata failed stability criteria and were omitted.

| Age Group | Code | Diagnosis Name | Same day OR (lag 0) |
|---|---|---|---|
| **18-24 years** | E86 | Volume depletion | 1.50 (1.24–1.81) |
|  | R21 | Rash and other nonspecific skin eruption | 1.39 (1.14–1.70) |

| Age group | Code | Diagnosis | OR (95% CI) |
|---|---|---|---|
| **25-44 years** | W34 | Accidental discharge and malfunction from other and unspecified firearms and guns | 1.80 (1.25–2.60) |
| | S40 | Superficial injury of shoulder and upper arm | 1.35 (1.09–1.68) |
| | S96 | Injury of muscle and tendon at ankle and foot level | 1.28 (1.08–1.52) |
| | S91 | Open wound of ankle, foot and toes | 1.27 (1.07–1.50) |
| | S90 | Superficial injury of ankle, foot and toes | 1.19 (1.01–1.40) |
| | E86 | Volume depletion | 1.15 (1.04–1.28) |
| **45–64 years** | T67 | Effects of heat and light | 4.57 (2.46–8.50) |
| | S92 | Fracture of foot and toe, except ankle | 1.38 (1.14–1.67) |
| | E86 | Volume depletion | 1.18 (1.08–1.29) |
| | R60 | Edema, not elsewhere classified | 1.11 (1.00–1.23) |
| **65 + years** | S92 | Fracture of foot and toe, except ankle | 1.42 (1.04–1.95) |
| | I95 | Hypotension | 1.13 (1.02–1.24) |

**Table 3. Heat-associated diagnoses by age group.** ICD-10 diagnosis codes with immediate (same day, lag-0) increases in emergency department visits during extreme heat, where the lower bound of the 95% confidence interval exceeded 1, stratified by age group. Codes are ordered by decreasing same day odds ratios (ORs).

In sex-stratified analyses, the most consistent finding across both sexes was an elevated odds of ED visits for ***Volume depletion*** (E86) during extreme heat (Table 4). Among females, additional elevations were evident for ***Effects of heat and light*** (T67) and ***Lower-extremity injuries***—open wounds of the ankle/foot/toes (S91) and muscle/tendon injury at the ankle/foot (S96)—as well as ***Hypotension*** (I95). Among males, increases were concentrated in fluid/renal outcomes, including ***Edema*** (R60) and ***Acute kidney failure*** (N17). Estimates for ***Exposure to excessive natural heat*** (X30) did not meet stability criteria in either females or males and estimates for ***Effects of heat and light*** (T67) did not meet stability criteria among males; therefore, these results are not tabulated.

| Sex | Code | Diagnosis Name | Same day OR (lag 0) |
|---|---|---|---|
| **Female** | T67 | Effects of heat and light | 4.37 (2.52–7.59) |
| | S91 | Open wound of ankle, foot and toes | 1.29 (1.09–1.53) |
| | S96 | Injury of muscle and tendon at ankle and foot level | 1.29 (1.11–1.49) |
| | I95 | Hypotension | 1.13 (1.03–1.24) |
| | E86 | Volume depletion | 1.07 (1.00–1.15) |
| **Male** | E86 | Volume depletion | 1.18 (1.09–1.27) |
| | R60 | Edema, not elsewhere classified | 1.11 (1.00–1.22) |
| | N17 | Acute kidney failure | 1.08 (1.01–1.15) |

**Table 4. Heat-associated diagnoses by sex.** ICD-10 diagnosis codes with immediate (same day, lag-0) increases in emergency department visits during extreme heat, where the lower bound of the 95% confidence interval exceeded 1, stratified by sex (Female, Male). Codes are ordered by decreasing same day odds ratios (ORs).

In race/ethnicity–stratified analyses (Table 5), ***Volume depletion*** (E86) and ***Assault by bodily force*** (Y04) were elevated in both Black or African American and White patients. ***Superficial injury of the ankle/foot/toes*** (S90) was elevated in Black or African American and those categorized as 'Other', and ***Open wound of the ankle/foot/toes*** (S91) was elevated in White and 'Other' patients. Group-specific patterns included, for Black or African American patients, increases in ***Hypotension*** (I95) and ***Open wound of the shoulder/upper arm*** (S41); for White patients, marked elevation for ***Effects of heat and light*** (T67), together with ***Superficial abdominal/lower-back injury*** (S30) and ***Acute kidney failure*** (N17). In the Asian stratum, signals were limited to ***Striking against/struck by other objects*** (W22); however, this subgroup represented a very small proportion of the cohort, yielding wide CIs.

| Race/Ethnicity | Code | Diagnosis Name | Same day OR (lag 0) |
|---|---|---|---|
| **Asian** | W22 | Striking against or struck by other objects | 4.67 (1.03–21.23) |
| **Black or** | S41 | Open wound of shoulder and upper arm | 1.36 (1.05–1.77) |
| **African American** | S90 | Superficial injury of ankle, foot and toes | 1.19 (1.03–1.39) |
| | Y04 | Assault by bodily force | 1.16 (1.03–1.32) |
| | I95 | Hypotension | 1.13 (1.02–1.24) |
| | E86 | Volume depletion | 1.10 (1.03–1.18) |
| **White** | T67 | Effects of heat and light | 6.24 (3.31–11.75) |
| | S30 | Superficial injury of abdomen, low back, pelvis and external genitals | 1.38 (1.02–1.87) |
| | Y04 | Assault by bodily force | 1.27 (1.02–1.60) |
| | S91 | Open wound of ankle, foot and toes | 1.20 (1.00–1.44) |
| | E86 | Volume depletion | 1.14 (1.04–1.25) |
| | N17 | Acute kidney failure | 1.11 (1.01–1.21) |
| **Other** | S91 | Open wound of ankle, foot and toes | 1.78 (1.22–2.61) |
| | S90 | Superficial injury of ankle, foot and toes | 1.38 (1.03–1.86) |

**Table 5. Heat-associated diagnoses by race/ethnicity.** ICD-10 diagnosis codes with immediate (same day, lag-0) increases in emergency department visits during extreme heat, where the lower bound of the 95% confidence interval exceeded 1, stratified by race/ethnicity. Codes are ordered by decreasing same day odds ratios (ORs).

The comprehensive subgroup analysis stratified by geographic region of residence (Central, Far South, North, Northwest, South, Southwest, West Side) is provided as Table S2. In region-stratified analyses, heat-specific diagnoses were most prominent on the West Side, where ***Exposure to excessive natural heat*** (X30) and ***Effects of heat and light*** (T67) showed the largest same day increases. Injury-related elevations were distributed across several regions: the Northwest Side exhibited higher odds of ***Open wound of the elbow/forearm*** (S51) and ***Superficial ankle/foot/toe injury*** (S90); the South Side showed increases in ***Fracture of the foot/toe*** (S92) and ***Superficial knee/lower-leg injury*** (S80); the Southwest Side had elevations in ***Accidental discharge and malfunction from firearms and guns*** (W34), ***Foot/toe fracture*** (S92), and ***Rash/nonspecific eruption*** (R21); and the West Side also had higher odds of ***Open wound of the ankle/foot/toes*** (S91) and ***Assault by bodily force*** (Y04). Medical diagnoses were elevated on the North Side, including ***Volume depletion*** (E86) and ***Edema*** (R60).

**Sensitivity Analysis**

We conducted four sensitivity analyses to assess the robustness of our findings (Table 6). Core associations were largely robust across alternative model specifications: ***Heat-specific diagnoses*** (T67, X30), ***Volume depletion*** (E86), ***Acute kidney failure*** (N17), and ***Open wound of ankle, foot and toes*** (S91). Changing the reference temperature from 50$^{th}$ to the 70$^{th}$ percentile (with extreme heat exposure fixed at the 95$^{th}$ percentile) removed ***Hypotension*** (I95), ***Injury of muscle/tendon at ankle/foot*** (S96), and ***Edema*** (R60), with no additional diagnoses detected. Using fixed, non-overlapping 28-day windows with three same-weekday controls preserved the overall pattern but dropped ***Injury of muscle/tendon at ankle/foot*** (S96), ***Accidental discharge and malfunction from other firearm discharge*** (W34), and ***Assault by bodily force*** (Y04); under this specification, ***Glomerular disorders*** (N08), ***Pedal-cycle driver injured in collision with a pedestrian/animal*** (V10), and ***Open wound of shoulder/upper arm*** (S41) were newly discovered. Re-parametrizing the exposure-response with a cubic B-spline with one internal knot at the 50$^{th}$ percentile eliminated ***Assault by bodily force*** (Y04), ***Accidental discharge and malfunction from other firearm discharge*** (W34), ***Injury of muscle/tendon at ankle/foot*** (S96), and ***Edema*** (R60), while ***Open wound of knee/lower leg*** (S81) newly appeared. Extending the lag window to 0–5 days with a natural-spline lag basis using two log-spaced internal knots produced similar results, with only ***Injury of muscle/tendon at ankle/foot*** (S96) no longer retained, and no additional diagnoses detected.

Overall, these sensitivity checks suggest that the strongest and most stable associations were for ***Heat-specific outcomes*** (T67, X30), ***Volume depletion*** (E86), ***Acute kidney failure*** (N17), and ***Open wounds of the ankle/foot/toes*** (S91). In contrast, ***Injury of muscle/tendon at***

*ankle/foot* (S96) was consistently sensitive to model specification, and sporadic single-detection codes (N08, V10, S41, S81) appear less robust. Taken together, the results reinforce the robustness of the primary findings while highlighting a subset of outcomes that warrant more cautious interpretation.

| Sensitivity Analysis | Dropped from primary | Newly detected (sensitivity-only) |
|---|---|---|
| (i) Higher reference temperature (P70; extreme heat = P95) | I95, S96, R60 | — |
| (ii) Controls = fixed 28-day, 3 same weekdays | S96, W34, Y04 | N08 (Glomerular disorders) V10 (pedal cycle driver injured in collision with a pedestrian/animal) S41 (Open wound of shoulder and upper arm Injury) |
| (iii) Cubic B-spline with one internal knot at P50 for same-day effect | R60, W34, Y04, S96 | S81 (Open wound of knee and lower leg) |
| (iv) Longer lag-windows (0-5 days) using two log-spaced lag knots | S96 | — |

**Table 6. Sensitivity analyses: Changes relative to the primary model.** Changes in ICD-10 codes across four sensitivity analyses relative to the primary time-stratified case-crossover distributed lag non-linear model (reference temperature = 50th percentile; extreme heat exposure = 95th percentile; calendar month × weekday strata; quadratic B-spline with one internal knot). 'Dropped from primary' indicates diagnosis codes that are no longer significant in the model, while 'Newly detected (sensitivity-only)' indicates diagnosis codes that became significant only under that sensitivity model specification.

## Discussion

We conducted an HWAS in Chicago to systematically identify diagnostic categories of ED visits that were associated with extreme heat across five major healthcare systems. Across more than 370,000 patients and 900,000 ED visits between May and September from 2011 to 2023, we systematically evaluated the full ICD-10 diagnostic spectrum within an integrated urban ED system using a two-stage analytic framework. Our approach firstly identified 38 diagnosis codes in the first-stage screening process and narrowed these to 11 diagnosis codes associated with extreme heat in the second-stage, spanning heat-related illnesses, fluid/electrolyte and renal disorders, injuries and poisoning, and other external causes of morbidity. These associations remained consistent across alternative model specifications, underscoring their stability. Moreover, subgroup analyses revealed heterogeneity in heat vulnerability by age, sex, race/ethnicity, and region of residence. For example, younger and working-age adults showed increased odds of volume depletion during extreme heat, whereas older adults exhibited elevated risks of hypotension. Such variation indicates that certain populations bear a disproportionate burden of heat-related morbidity, underscoring that prevention strategies must be targeted to specific populations.

Our study provides a comprehensive characterization of heat-associated morbidity that both reinforces and advances the existing literature. Excessive heat exposure has been shown to be associated with heat-related illness (18, 19) and broad diagnostic categories, including renal, cardiovascular, respiratory, electrolyte, mental disorders (20-24, 26-28, 47), and injury-related diagnoses (7, 29). Much of this literature has focused on older adults (7, 10, 12, 30, 34), and to a lesser extent, pediatric populations, who are widely recognized as particularly susceptible to heat (31, 48, 49). However, to our knowledge, no studies have systematically assessed individual-level diagnosis codes from ED records across the broader adult population.

Consistent with earlier studies, we observed substantial elevations in ED visits for narrowly-defined heat-related illness codes (18, 19). Although prior studies have frequently highlighted cardiovascular outcomes (19-22, 24, 25, 47), our analysis did not identify significant short-term increases in cardiovascular-related ED visits. Our findings suggest that, in ED settings, the burden of extreme heat is reflected not only through circulatory and fluid-related complications such as hypotension, edema, and volume depletion, but also through acute kidney failure. Taken together, these outcomes indicate that heat stress is associated with a broader spectrum of physiologic disturbances—spanning fluid balance and renal diseases—rather than being limited to cardiovascular or respiratory diseases.

Our study also contributes to the growing recognition of trauma-related clinical diagnoses—including injuries (S91, S96), firearm-related incidents (W34), and assault-related conditions (Y04) and their association with heat-sensitive outcomes. There have been studies that demonstrated higher risks of accidental injuries during hot days, which support our findings (7, 29). The identification of assault-related codes further reinforces evidence that extreme heat is associated with clinical diagnoses related to trauma and assault, which may support related behavioral health research linking elevated temperatures to heightened aggression and psychiatric distress (50, 51). Although psychiatric diagnoses did not appear among our identified codes, the external-cause signals may indirectly reflect heat's effects on mental and behavioral health that are not fully captured in clinical diagnostic coding (50-52).

Sensitivity analyses confirmed that identified associations were robust to alternative modeling choices, including redefining the reference temperature, altering the control-day selection window, varying spline placement, and substituting alternative spline bases. Although the effect magnitudes shifted modestly under these specifications, the overall pattern of heat-sensitive diagnoses remained stable, reinforcing the reliability of our findings.

Together, we have demonstrated that our Chicago-based HWAS not only reproduces several well-established associations but also advances prior work through its comprehensive and robust design, which systematically evaluated all available ICD-10 codes across the study population rather than relying on a pre-specified subset. Our stratified analyses further show that these vulnerabilities are not limited to older adults but also affect working-age populations, consistent with emerging evidence that occupational exposures increase susceptibility among working-age individuals during heat events (53).

Despite our study's strengths, several limitations should be acknowledged. First, the study population was unevenly distributed across demographic and geographic strata, with certain demographic (such as Asians) and geographic (such as Far South Side) groups underrepresented. This imbalance may have constrained the precision of subgroup analyses and limits the generalizability of our findings to populations that are not well represented in the dataset. Second, the analytic cohort was restricted to adults; pediatric patients were not included, although prior studies suggest that children are particularly vulnerable to extreme heat (31, 48, 49). Third, for several heat-specific diagnoses (e.g., X30 and T67), the lag-0 ORs were elevated but accompanied by wide 95% CIs, reflecting imprecision due to the limited number of events for these diagnosis codes in the dataset—which further underscores the challenge of defining heat-related health outcomes given their diverse presentation, especially in emergency settings. This limitation was further exacerbated in subgroup analyses, where event counts were even smaller. This limitation is inherent to our analytic design, which evaluates risks at the level of individual diagnosis codes rather than aggregated categories. Fourth, our analysis was restricted to patients who utilized healthcare services in 5 healthcare systems in Chicago, a dense urban environment. Accordingly, the extent to which our results are transferable to suburban or rural contexts remains uncertain, and future work should examine potential geographical differences. Fifth, we relied on cross-sectional ED visit data rather than longitudinal patient records, precluding the ability to track cumulative heat impacts, recurrent hospitalizations, or the modifying influence of long-term health conditions and medication use. Sixth, socioeconomic indicators such as income, education, and housing quality were not incorporated. Their omission raises the potential for residual confounding and may obscure structural inequities in heat-related health risks. Finally, our study links patient residential census tract to temperature with the assumption that this best represents a patient's exposure. This approach does not account for movement within the city or environmental settings, exposure to extreme heat where the location of occupational work is different from the location of residency such as time spent at work, use of air conditioning, or time indoors during heat events.

In summary, our large-scale citywide HWAS shows that extreme heat is associated with increased ED presentations beyond classical heat illness, including renal dysfunction and fluid and electrolyte disturbances, as well as injury- and assault-related diagnoses, with heterogeneity by age, sex, race/ethnicity, and geographic region. By presenting a data-driven and comprehensive characterization of heat-sensitive conditions, this study can be used to more completely define heat-related health outcomes that enable targeted public health interventions and strengthen clinical preparedness for future extreme heat events.

## Materials and Methods

### Data

We used ED encounters captured in CAPriCORN (43) EHR data from 5 healthcare systems to construct a Chicago citywide cohort of ED encounters among adult patients (916,904 ED visits; 372,140 unique patients). Inclusion criteria comprised of ED visits by adults (aged ≥18 years) who sought care at a CAPriCORN-affiliated institution, with geocoded residence located within Cook County, Illinois (the county that includes the City of Chicago), during the warm-season (May through September) of 2011–2023, and with valid ICD-10 diagnostic codes. Because our study period included the transition from ICD-9 to ICD-10 codes (mandated in the United States [U.S.] by October 1, 2015), to standardize the diagnoses we mapped the ICD-9 codes to their corresponding ICD-10 codes using the Centers for Medicare and Medicaid Services General Equivalence Mappings guidance and removed the duplicate codes within each ED encounter. This process yielded a final set of ICD-10 diagnosis codes (n = 1,803 distinct categories) extracted from the EHRs. We chose ICD-10 (54) as the analytic standard because it provides more granular and internationally recognized diagnostic classifications, which improves consistency across sites and time periods. Although ICD codes are primarily designed for billing in the U.S., they are widely leveraged in epidemiological and health services research.

Patient characteristics were stratified by sex (Female, Male, and Other), age group (18-24, 25-44, 45-64, ≥ 65 years), and race/ethnicity (Asian, Black or African American, White, and Other). These categorizations followed the standardized demographic classifications used within CAPriCORN (43). For race/ethnicity, groups with relatively small representation–American Indian or Alaska Native (0.6%), Native Hawaiian or Other Pacific Islander (0.2%), and individuals reporting multiple races (0.3%) were aggregated into the 'Other' category. For geographic region of residence, we stratified the region into seven Chicago Department of Public Health (CDPH) planning regions—Central, North, Northwest, South, Southwest, Far South, and West Side—based on census tract centroid geography (55). A visualization of the region stratification is provided in the Supplementary Materials (Figure S5).

For exposure assessment, daily temperature data were sourced from Daymet (56) and linked to census tract of each patient's residential address at the date of their ED encounter(s). Estimates for daily maximum air temperature (°C) from Daymet 1 km$^2$ grid cells that fell within a patient residential census tract were averaged for each ED encounter day, per patient. Extreme heat was defined as days on which the Daymet-derived daily maximum temperature exceeded the 95th percentile of the warm-season (May–September) distribution (33.67°C), consistent with definitions commonly applied in prior studies (19, 57, 58).

### Statistical Analysis

To comprehensively assess the health impacts of extreme heat, we applied an HWAS design. This approach systematically screens across all diagnosis codes for an extreme heat association, rather than restricting analyses to a limited set of pre-specified outcomes, enabling more complete capture and discovery of heat-sensitive conditions. Given the large number of initial ED visit diagnosis codes, we adopted a two-stage analytic strategy for the HWAS. For the first stage, we conducted a screening procedure using quasi-Poisson regression to identify temperature-sensitive diagnoses, taking into account the over-dispersed daily ED visit counts by diagnosis code. For each ICD-10 code, the daily count of ED visits with that code served as the outcome and citywide daily maximum temperature as the exposure, while controlling for year, month, day of week, and federal holidays for long-term trends and temporal confounding. We retained diagnosis codes that met the following criteria: (1) the temperature coefficient was positive and statistically significant after BH adjustment (45); (2) at least 30% of ED visits for that diagnosis occurred on days when the daily maximum temperature exceeded the 70$^{th}$ percentile threshold (29.74°C) of the overall daily maximum temperature distribution during the warm-season; and (3) at least 100 counts of the diagnosis code were recorded during the warm-season in the ED systems. We used the 70th percentile (29.74°C) of the warm-season temperature distribution to capture commonly experienced, rather than rare temperature extremes, for the first screening stage. In parallel, we required that $\geq$ 30% of ED visits for a given code occurred on $\geq$ 70$^{th}$ percentile temperature days, as a support criterion that ensures sufficient observations under warm conditions and limits signals driven by a few extreme heat days. Requiring diagnosis codes to have a frequency $\geq$ 100 ensured that our first screening stage did not capture codes that are rarely used in clinical practice.

In the second stage, the resulting candidate diagnoses, identified from the first screening stage, were analyzed individually using DLNMs within a time-stratified case-crossover design (13), with strata defined by calendar month and year, and control days matched to the case days within the same stratum. In the case-crossover framework, each ED visit served as its own case day, with control days sampled from the same day of week of the same calendar month to account for seasonal and long-term temporal trends. This design inherently controls for time-invariant individual- and area-level confounders. Daily maximum temperature was modeled with a DLNM cross-basis to capture both the non-linear exposure–response and distributed lag effects. The exposure dimension used a quadratic B-spline with a single internal knot at the 50th percentile of the warm-season temperature distribution, and the lag dimension used a natural spline over lags 0–3 days with a single internal knot placed on the log-transformed lag scale, concentrating flexibility at shorter time lags, typical of heat-related ED presentations (19, 59, 60). Conditional logistic regression, additionally controlling for federal holidays, was used to estimate both lag-specific and cumulative ORs for extreme heat (95$^{th}$ percentile of the warm-season temperature distribution, 33.67°C) compared with moderate temperature (50$^{th}$ percentile of the warm-season temperature distribution, 27.59°C). Diagnosis codes with a lower bound of the 95% CI at lag 0 greater than 1 were considered to have a statistically significant immediate extreme heat effect ($\alpha = 0.05$).

To evaluate potential effect modification by demographic and geographic factors, we performed stratified analyses retaining the same DLNM specification. Cumulative and lag-specific ORs were estimated separately within strata defined by patient sex (Female, Male, and Other), age group (18–24, 25–44, 45–64, and $\geq$ 65 years), race/ethnicity (Asian, Black or African American, White, and Other), and geographic region of residence (Central, North, Northwest, South, Southwest, Far South, and West Side Chicago). Within each stratum, we refitted the DLNM within the time-stratified case-crossover framework to obtain stratum-specific point estimates and standard errors. In subgroup analyses, we prespecified a stability rule to address sparse events. A diagnosis-stratum was flagged as unstable if any of the following occurred: model non-convergence or quasi-separation (non-finite coefficients/variance or maximum iterations reached), unusually large uncertainty in cross-basis coefficients (standard error > 10), or exp(β) non-finite or > 100. Diagnosis-stratum flagged as unstable were omitted from the tables and not used for inference, while other diagnoses within the same stratum were retained.

We conducted a series of sensitivity analyses to assess the robustness of our findings to key modeling decisions. First, we varied the reference temperature for extreme heat comparisons by replacing the 50$^{th}$ percentile (27.59°C) with the 70$^{th}$ percentile (29.74°C) of the warm-season maximum temperature distribution, while retaining the 95$^{th}$ percentile as the threshold for defining extreme heat. Second, we evaluated the impact of the control day selection strategy by employing fixed, non-overlapping 28-day referent windows in place of strata defined by calendar month. Third, we estimated the same-day effect with a cubic B-spline exposure basis with one internal knot at the 50$^{th}$ percentile. Finally, we extended the lag window to 0-5 days and fitted a natural-spline lag basis with two log-spaced internal knots. For all of these sensitivity analyses, we compared resulting heat-related diagnosis codes to those identified in our primarily model specification.

**Acknowledgments**

**Funding:**
American Heart Association (AHA) Predoctoral Fellowship 24PRE1193628 (PMG)
Northwestern University's Buffett Institute for Global Affairs Defusing Disasters working group.

**Author contributions:**
    Conceptualization: HJ, PMG, BWB, DEH, JLC, ANK
    Methodology: HJ, PMG, BWB
    Software: HJ, PMG, BWB
    Validation: HJ, PMG, BWB, DEH, JLC, ANK
    Formal analysis: HJ
    Investigation: HJ, PMG, BWB
    Data curation: PMG, BWB, DEH
    Visualization: HJ, PMG, BWB
    Writing—original draft: HJ
    Writing—review & editing: HJ, PMG, BWB, DEH, JLC, ANK
    Supervision: PMG, BWB, ANK
    Project administration: PMG, BWB, ANK
    Funding acquisition: ANK, PMG, DEH
    Resources: PMG, BWB, ANK

**Competing interests:** Authors declare that they have no competing interests.

**Data and materials availability:** The findings reported in this study were enabled through a collaboration with the Chicago Area Patient-Centered Outcomes Research Network (CAPriCORN). CAPriCORN is a partnership between healthcare and research institutions that provides data through a federated harmonized common data model and works jointly with a Patient Community Advisory Committee, community-based organizations (CBOs), and non-profit organizations committed to enabling and delivering patient-centered clinical research and public health projects. We acknowledge CAPriCORN's partners, the Chicago Area Institutional Review Board (CHAIRb), which serves as the central IRB of record for CAPriCORN-supported research, and the Medical Research Analytics and Informatics Alliance (MRAIA), which serves as the network's honest data broker.

# Supplementary Materials

**This PDF file includes:**
    Figs. S1 to S5
    Tables S1 to S2

**Fig. S1.**

Age distributions of emergency department visits for diagnosis codes retained in the second-stage screening step, stratified by non-extreme (< 95th percentile) and extreme (≥ 95th percentile) heat days.

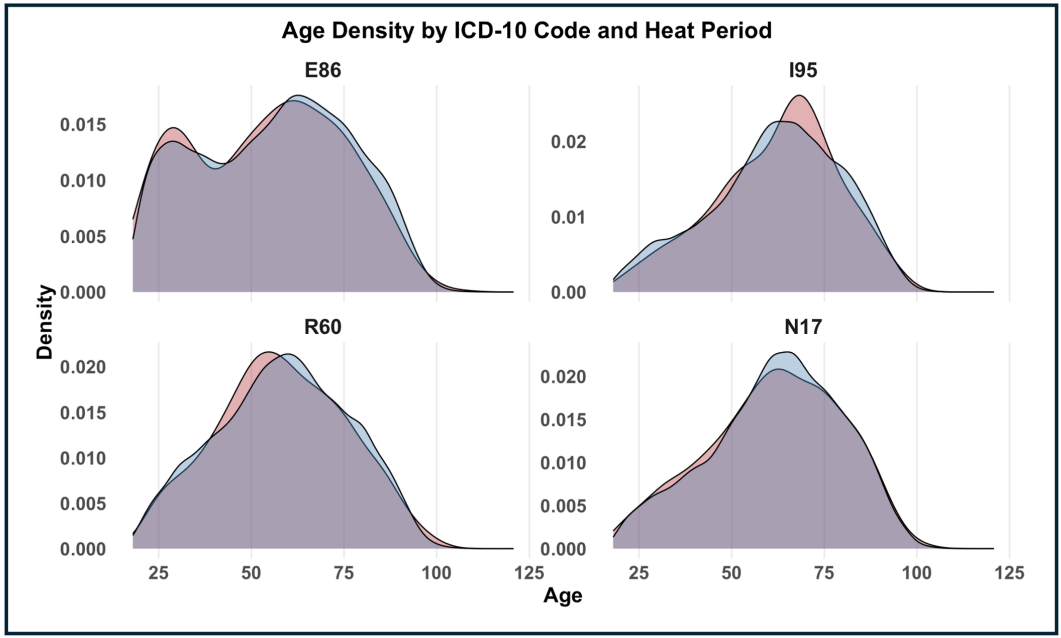

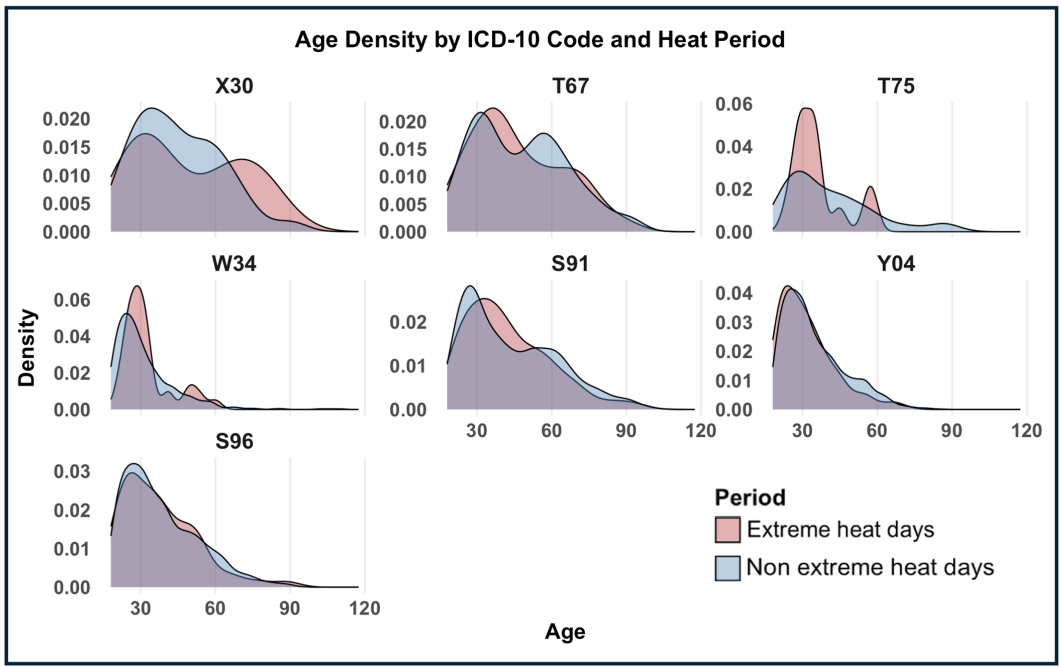

**Fig. S2.**

Sex distributions of emergency department visits for diagnosis codes retained in the second-stage screening step, stratified by non-extreme (< 95th percentile) and extreme (≥ 95th percentile) heat days.

(a) Disease related Diagnosis

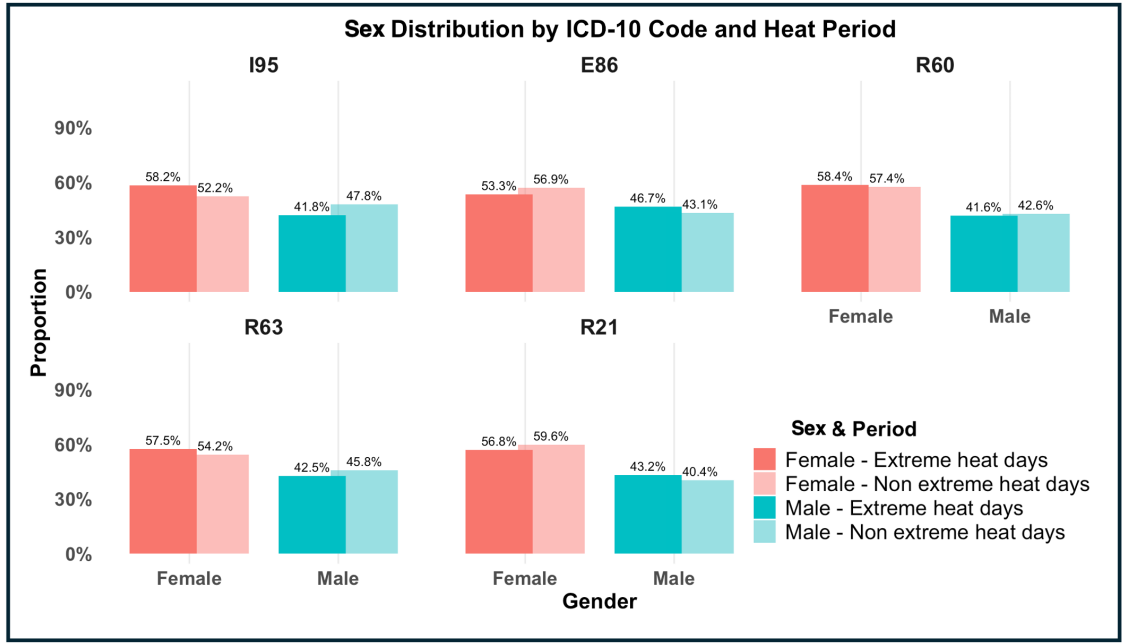

(b) Injury/Poisoning/External Causes related Diagnosis

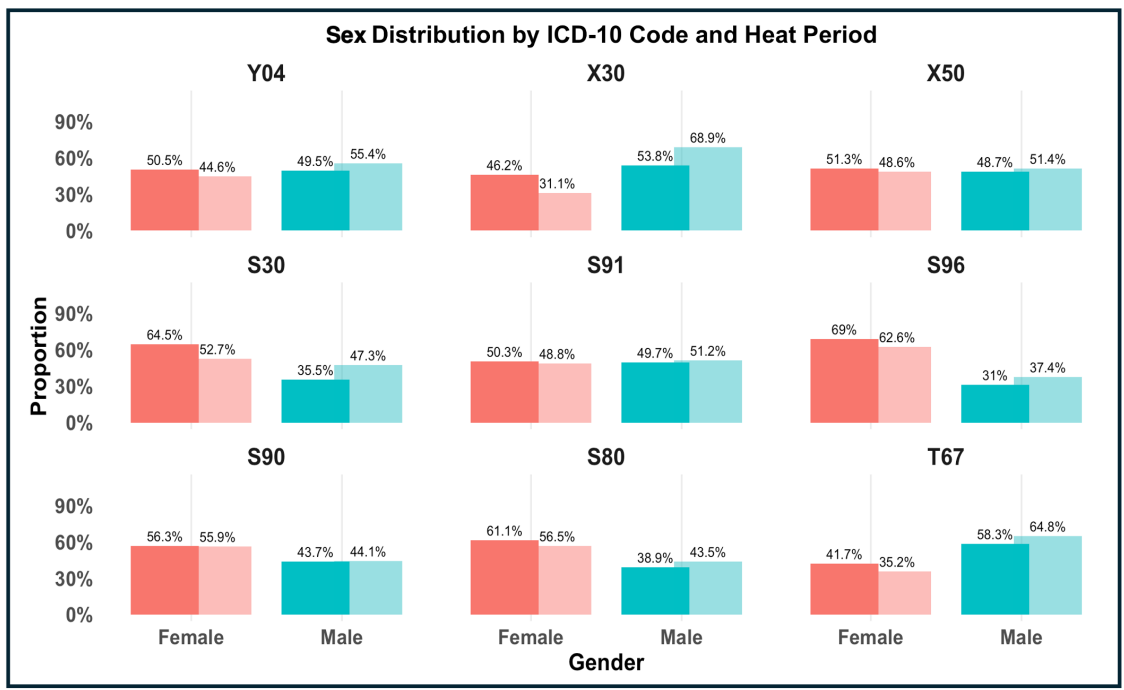

**Fig. S3.**

Race/ethnicity distributions of emergency department visits for diagnosis codes retained in the second-stage screening step, stratified by non-extreme (< 95th percentile) and extreme (≥ 95th percentile) heat days.

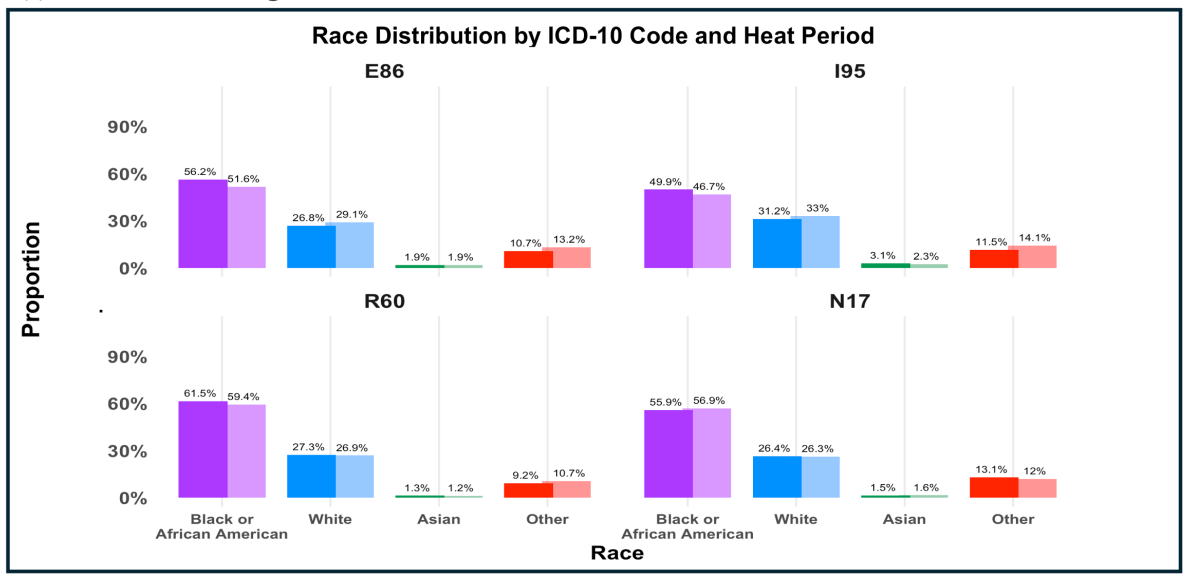

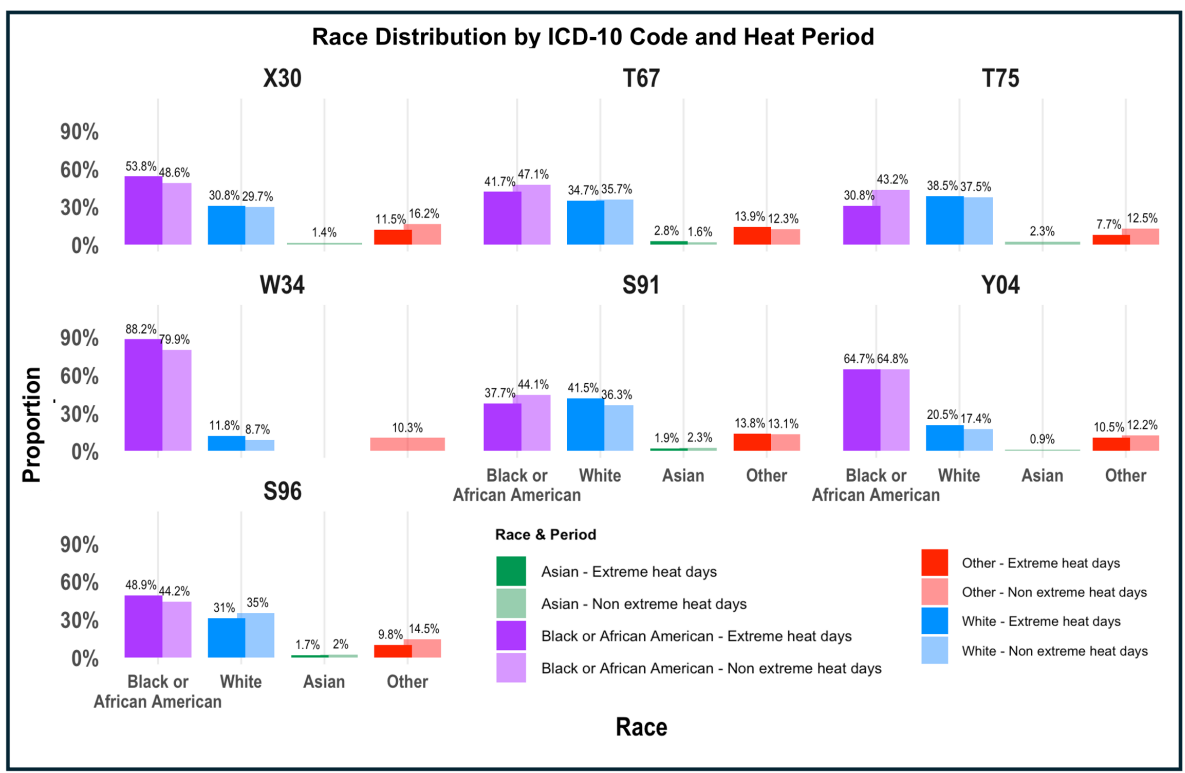

**Fig. S4.**

Geographic distributions of emergency department visits by heat exposure. Regional residence patterns (Central, North, Northwest, South, Southwest, Far South, West Side) for diagnosis codes retained in the second-stage screening step, stratified by non-extreme (< 95th percentile) and extreme (≥ 95th percentile) heat days.

**(a) Disease related Diagnosis**

Relative Frequency of ED Visits by Tract in Chicago
Proportion of ED Visits by Region and Heat Period

## E86

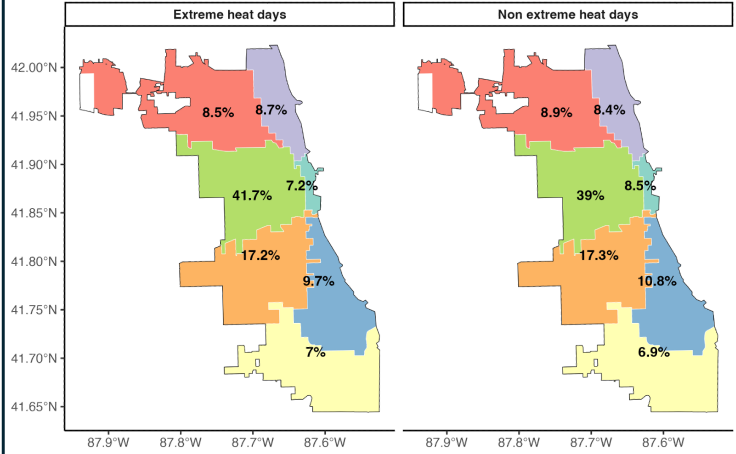

## I95

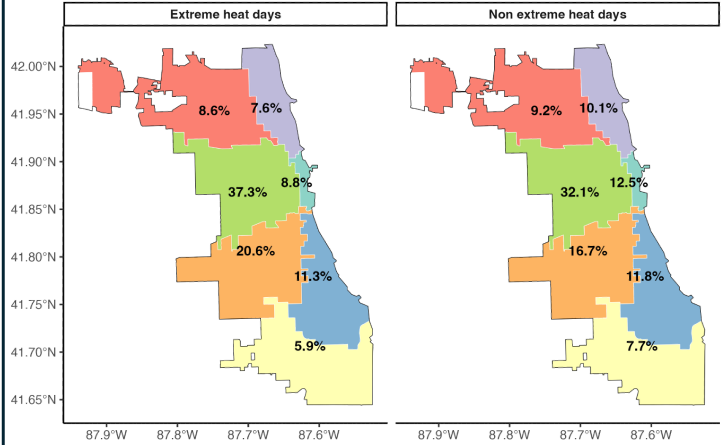

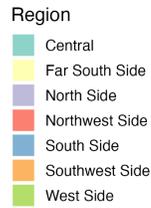

## R60

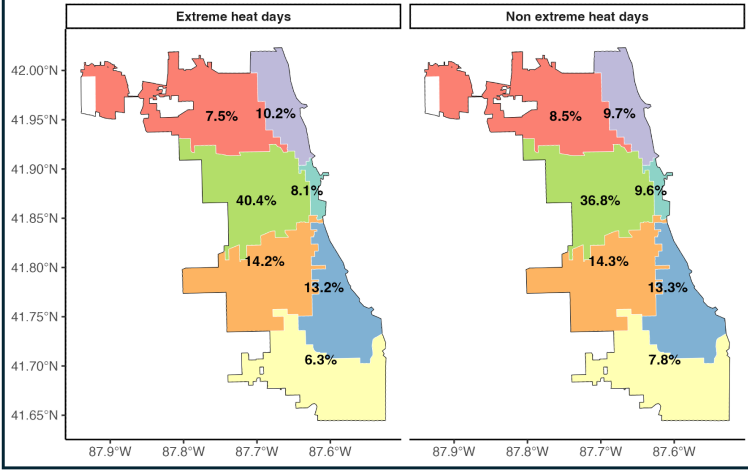

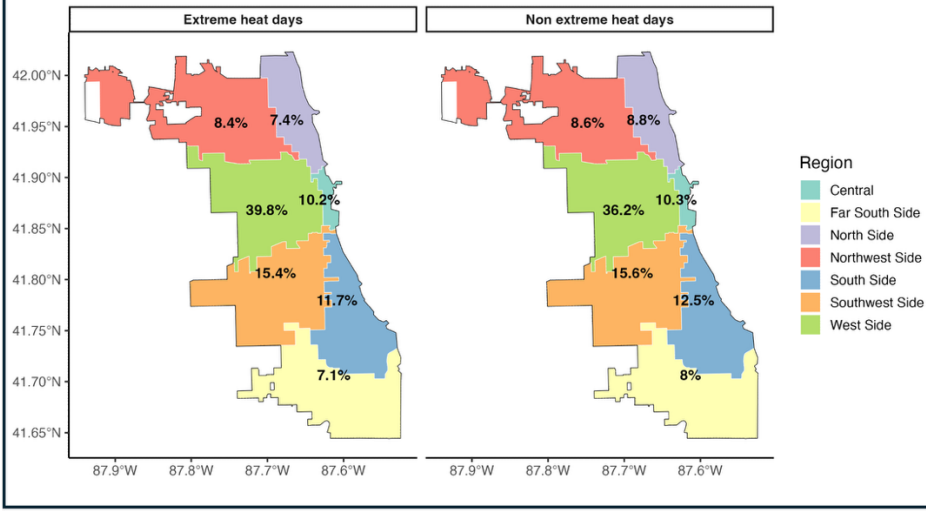

(a) Disease related Diagnosis

Relative Frequency of ED Visits by Tract in Chicago
Proportion of ED Visits by Region and Heat Period

## (b) Injury/Poisoning/External Causes related Diagnosis

Relative Frequency of ED Visits by Tract in Chicago
Proportion of ED Visits by Region and Heat Period

### X30

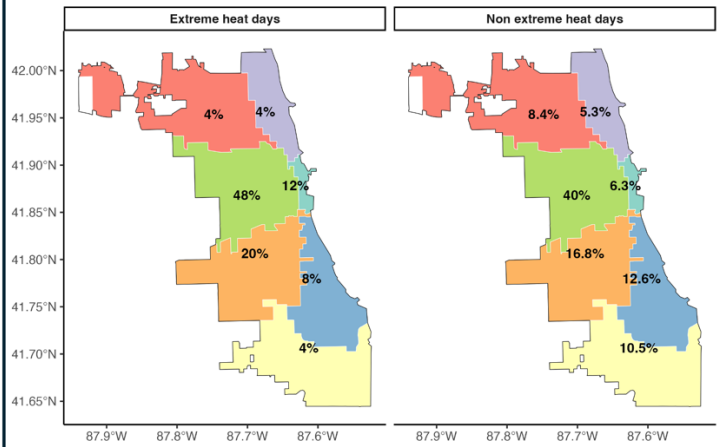

### T67

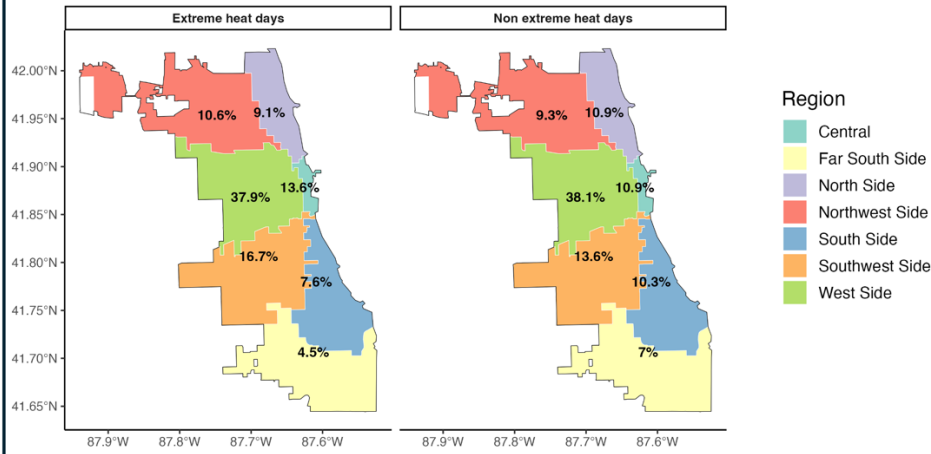

### T75

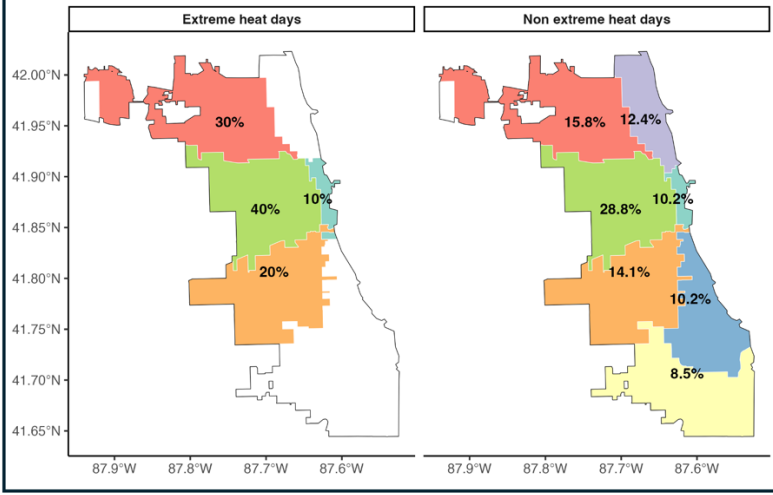

## (b) Injury/Poisoning/External Causes related Diagnosis

Relative Frequency of ED Visits by Tract in Chicago
Proportion of ED Visits by Region and Heat Period

### W34

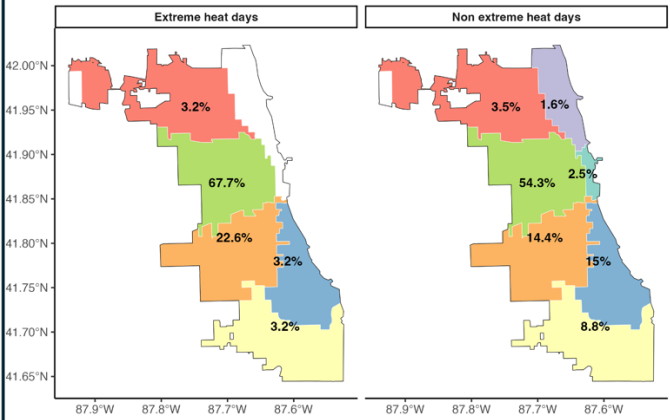

### S91

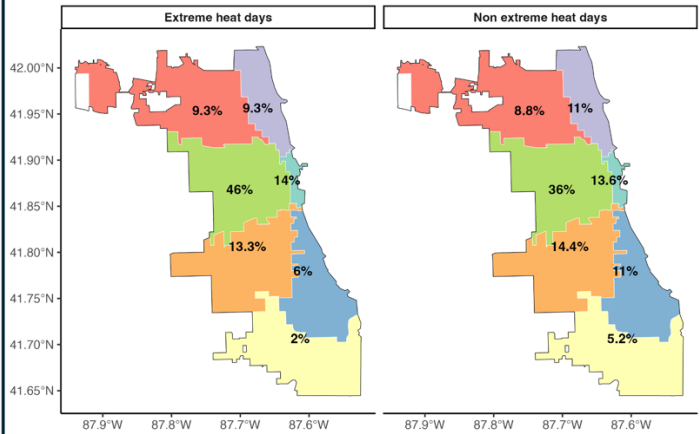

### Y04

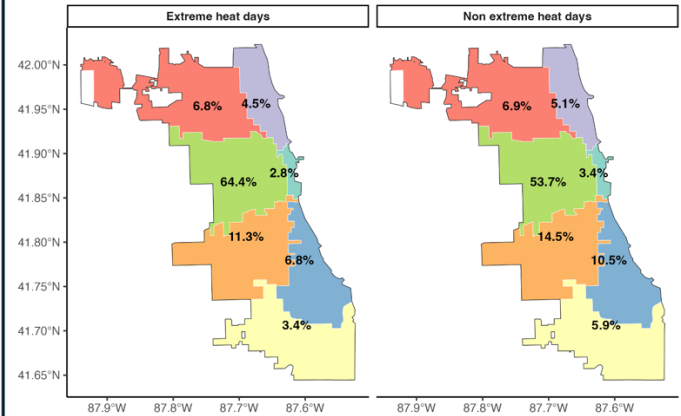

**(b) Injury/Poisoning/External Causes related Diagnosis**

Relative Frequency of ED Visits by Tract in Chicago
Proportion of ED Visits by Region and Heat Period

S96

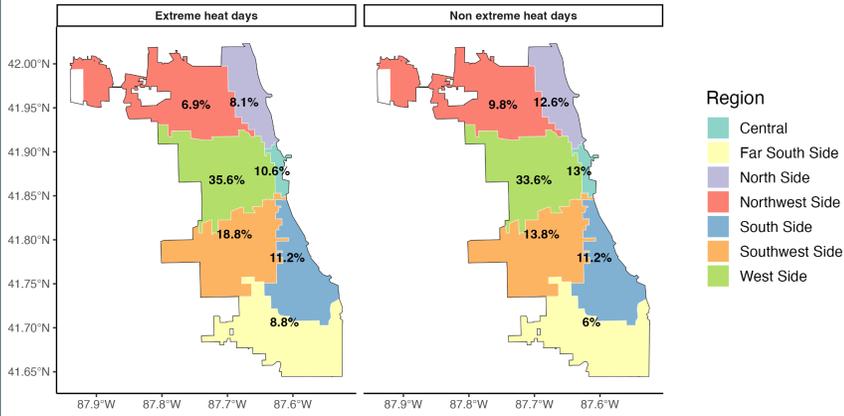

**Fig. S5.**

Geographic stratification of Chicago into seven regions as defined by the Chicago Department of Public Health (CDPH) Planning Regions: Central, North, Northwest, South, Southwest, Far South, and West Sides.

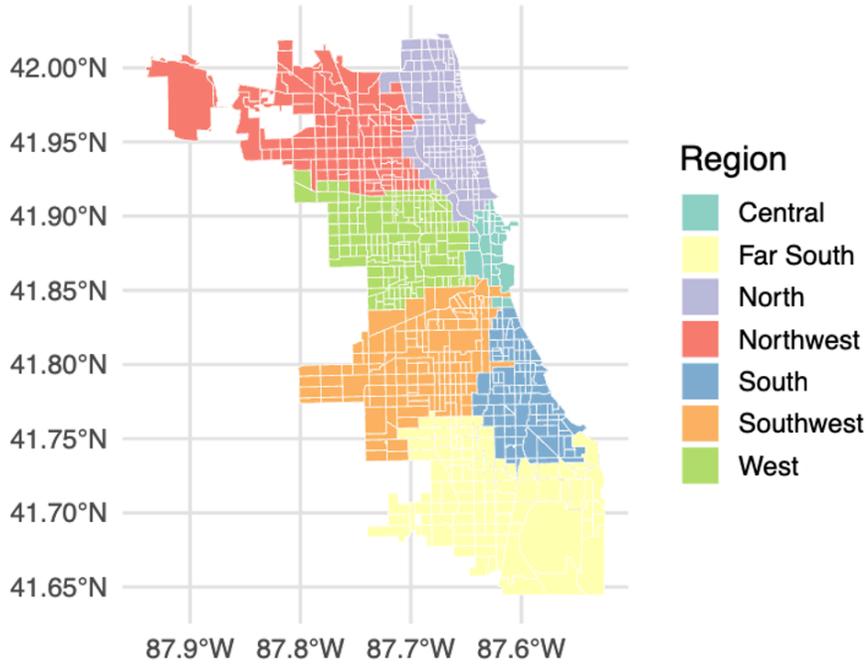

**Table S1.**

First-stage screening results for heat-sensitive emergency department diagnoses. ICD-10 codes with statistically significant positive associations with daily maximum temperature, adjusted by Benjamini–Hochberg false discovery rate, occurring in ≥ 30% of emergency department visits on days exceeding the 70th percentile (29.74°C) of the warm-season (May-September) daily maximum temperature distribution, and frequency ≥ 100 of the diagnosis code recorded during the warm-season. Columns report ICD-10 code, description, incidence rate ratio (IRR), 95% confidence interval (lower, upper), adjusted P value, and relative frequency.

| Code | Description | IRR | Lower | Upper | Adj.p | Rel.freq |
|---|---|---|---|---|---|---|
| X30 | Exposure to excessive natural heat | 1.33 | 1.25 | 1.42 | ≤ 0.01 | 0.78 |
| T67 | Effects of heat and light | 1.32 | 1.26 | 1.38 | ≤ 0.01 | 0.76 |
| T75 | Other and unspecified effects of other external causes | 1.07 | 1.03 | 1.11 | 0.01 | 0.36 |
| T24 | Burn and corrosion of lower limb, except ankle and foot | 1.04 | 1.01 | 1.07 | 0.02 | 0.35 |
| T22 | Burn and corrosion of shldr/up lmb, except wrist and hand | 1.03 | 1.01 | 1.06 | 0.03 | 0.34 |
| V19 | Family history of other conditions | 1.03 | 1.02 | 1.05 | ≤ 0.01 | 0.35 |
| T63 | Toxic effect of contact with venomous animals and plants | 1.03 | 1.01 | 1.05 | 0.04 | 0.31 |
| W25 | Contact with sharp glass | 1.03 | 1.01 | 1.04 | 0.01 | 0.35 |
| W34 | Acc disch and malfunct from oth and unsp firearms and guns | 1.03 | 1.01 | 1.05 | 0.04 | 0.36 |
| W50 | Acc hit, strk, kick, twist, bite or scratch by another prsn | 1.03 | 1.01 | 1.04 | 0.04 | 0.35 |
| S91 | Open wound of ankle, foot and toes | 1.02 | 1.02 | 1.03 | ≤ 0.01 | 0.33 |
| X99 | Assault by sharp object | 1.02 | 1.01 | 1.04 | 0.04 | 0.31 |
| S71 | Open wound of hip and thigh | 1.02 | 1.01 | 1.04 | 0.02 | 0.32 |
| N08 | Glomerular disorders in diseases classified elsewhere | 1.02 | 1.01 | 1.04 | 0.01 | 0.32 |
| S41 | Open wound of shoulder and upper arm | 1.02 | 1.01 | 1.04 | 0.01 | 0.34 |
| S51 | Open wound of elbow and forearm | 1.02 | 1.01 | 1.03 | ≤ 0.01 | 0.33 |
| W57 | Bit/stung by nonvenom insect and oth nonvenomous arthropods | 1.02 | 1.01 | 1.03 | 0.01 | 0.33 |
| S50 | Superficial injury of elbow and forearm | 1.02 | 1.01 | 1.03 | ≤ 0.01 | 0.33 |
| E36 | Intraoperative complications of the endocrine system | 1.02 | 1.01 | 1.03 | ≤ 0.01 | 0.31 |
| V10 | Personal history of malignant neoplasm | 1.02 | 1.01 | 1.04 | 0.04 | 0.31 |
| S92 | Fracture of foot and toe, except ankle | 1.02 | 1.01 | 1.03 | ≤ 0.01 | 0.31 |
| S90 | Superficial injury of ankle, foot and toes | 1.02 | 1.01 | 1.03 | ≤ 0.01 | 0.32 |
| S30 | Superfic inj abdomen, low back, pelvis and external genitals | 1.02 | 1.01 | 1.03 | 0.01 | 0.32 |
| S96 | Injury of muscle and tendon at ankle and foot level | 1.02 | 1.01 | 1.02 | 0.01 | 0.30 |
| S40 | Superficial injury of shoulder and upper arm | 1.01 | 1.01 | 1.02 | 0.02 | 0.33 |
| S81 | Open wound of knee and lower leg | 1.01 | 1.00 | 1.02 | 0.03 | 0.31 |
| R60 | Edema, not elsewhere classified | 1.01 | 1.01 | 1.02 | ≤ 0.01 | 0.30 |
| W22 | Striking against or struck by other objects | 1.01 | 1.01 | 1.02 | 0.01 | 0.33 |
| S52 | Fracture of forearm | 1.01 | 1.00 | 1.02 | 0.02 | 0.30 |
| S80 | Superficial injury of knee and lower leg | 1.01 | 1.01 | 1.02 | ≤ 0.01 | 0.32 |
| S62 | Fracture at wrist and hand level | 1.01 | 1.00 | 1.02 | 0.02 | 0.30 |
| Y04 | Assault by bodily force | 1.01 | 1.00 | 1.02 | 0.02 | 0.31 |
| E86 | Volume depletion | 1.01 | 1.01 | 1.01 | ≤ 0.01 | 0.31 |
| R21 | Rash and other nonspecific skin eruption | 1.01 | 1.00 | 1.01 | ≤ 0.01 | 0.30 |
| N17 | Acute kidney failure | 1.01 | 1.01 | 1.01 | ≤ 0.01 | 0.30 |
| R63 | Symptoms and signs concerning food and fluid intake | 1.01 | 1.00 | 1.01 | 0.02 | 0.31 |
| X50 | Overexertion and strenuous or repetitive movements | 1.01 | 1.00 | 1.01 | 0.04 | 0.30 |
| I95 | Hypotension | 1.01 | 1.00 | 1.01 | 0.02 | 0.30 |

**Table S2.**

Heat-associated diagnoses from second-stage screening stratified by geographic region of residence. ICD-10 codes with statistically significant immediate (same day, lag 0) heat-associated increases in emergency department visits, reported separately for each Chicago region (Central,

Far South, North, Northwest, South, Southwest, West Side). Results for the Central and Far South regions are excluded due to model non-convergence. Within each region, codes are ordered by decreasing same day odds ratios (ORs) for each region.

| Geographic Region | Code | Diagnosis Name | Same day OR (lag 0) |
|---|---|---|---|
| North Side | E86 | Volume depletion | 1.34 (1.11–1.62) |
| | R60 | Edema, not elsewhere classified | 1.26 (1.02–1.55) |
| Northwest Side | S51 | Open wound of elbow and forearm | 2.07 (1.05–4.08) |
| | S90 | Superficial injury of ankle, foot and toes | 1.53 (1.06–2.21) |
| South Side | S92 | Fracture of foot and toe, except ankle | 1.68 (1.12–2.52) |
| | S80 | Superficial injury of knee and lower leg | 1.37 (1.03–1.83) |
| Southwest Side | W34 | Accidental discharge and malfunction from other and unspecified firearms and guns | 2.11 (1.06–4.18) |
| | S92 | Fracture of foot and toe, except ankle | 1.46 (1.05–2.04) |
| | R21 | Rash and other nonspecific skin eruption | 1.23 (1.04–1.46) |
| West Side | X30 | Exposure to excessive natural heat | 9.48 (2.35–38.22) |
| | T67 | Effects of heat and light | 7.14 (3.88–13.14) |
| | S91 | Open wound of ankle, foot and toes | 1.22 (1.02–1.47) |
| | Y04 | Assault by bodily force | 1.19 (1.03–1.38) |